\documentstyle[prb,aps,twocolumn,epsfig]{revtex}

\def\prb{\it Phys. Rev. B}
\def\prl{\it Phys. Rev. Lett.}

\def\be{\begin{equation}}
\def\ee{\end{equation}}
\def\ba{\begin{eqnarray}}
\def\ea{\end{eqnarray}}

\def\LSCO{La$_{2-x}$Sr$_x$CuO$_4$}
\def\YBCO{YBa$_2$Cu$_3$O$_{7-\delta}$}

\def\C60{A$_x$C$_{60}$}
\def\LNSCO{La$_{1.6-x}$Nd$_{0.4}$Sr$_x$CuO$_{4}$}

\def\hty{high temperature superconductivity}
\def\hts{high temperature superconductors}

\begin{document}

\twocolumn[\hsize\textwidth\columnwidth\hsize\csname@twocolumnfalse\endcsname

\title
{Electronic Liquid Crystal Phases of a Doped Mott Insulator}

\author{S.~A.~Kivelson$^{a}$, E.~Fradkin$^{b}$, V.~J.~Emery$^{c}$}
\address{Dept. of Physics, U.C.L.A., Los Angeles, CA  90095$^{a}$, 
Dept. of Physics, University of Illinois, Urbana, IL 61801-3080$^{b}$, and 
Brookhaven National Laboratory, Upton NY 11973-5000$^{c}$}.
\date{\today}
\maketitle

]

\narrowtext

{\bf The character of the ground state of an antiferromagnetic insulator 
is fundamentally altered upon addition of even a small amount of
charge\cite{topo}.  The  added charges 
agglomerate along  domain walls at which the spin correlations, which may or
may not remain long-ranged, suffer a $\pi$ phase shift.
In two dimensions, these domain walls are ``stripes'' which are either 
insulating \cite{insul},
or conducting\cite{metal}, {\it i.e.} metallic rivers with their own low energy 
degrees of freedom. However, quasi one-dimensional metals \cite{1d} typically 
undergo a transition to an insulating ordered charge density 
wave (CDW) state at low temperatures. Here it is shown that such a transition 
is eliminated if the zero-point energy of transverse stripe fluctuations
is sufficiently large in comparison to the CDW coupling between stripes. 
As a consequence, there
exist novel, liquid-crystalline low-temperature phases -- an electron smectic, 
with crystalline order in one direction, but liquid-like correlations in 
the other, and an electron nematic with orientational order but no long-range
positional order. These phases, which constitute new states of matter, can be 
either {\hts} or two-dimensional anisotropic ``metallic'' non-Fermi liquids.
Evidence for the new phases may already have been obtained
by neutron scattering experiments in the cuprate superconductor, 
{\LNSCO}}.

A single metallic stripe is a prototypical example of the
one-dimensional  electron gas in an active 
environment\cite{spingap,active} about
which much is known.  In the absence of coupling 
between the stripes, this system is generally ``quantum critical'',  {\it i.e.} 
it
exhibits power law correlations and has an infinite  correlation length at 
temperature $T=0$.
Frequently it has a gap in its spin 
excitation
spectrum, $\Delta_s$, so the only low-energy degrees of freedom are the CDW
and the dual superconducting fluctuations whose susceptibilities 
diverge as $T\rightarrow 0$, as
\be
\chi_{CDW} \sim \Delta_s T^{-(2-K_c)}; 
\ \ \chi_{SC} \sim
\Delta_s T^{-(2-1/K_c)}
\label{eq:chi}
\ee
where $K_c$ is a non-universal
critical exponent\cite{emery} which depends on the magnitude and sign of the
interactions and satisfies $0 < K_c < 1$ for repulsive 
interactions. 

Direct evidence\cite{tranq,yamada,mook} of stripe
correlations in the cuprate superconductors, themselves doped antiferromagnets, 
has been
obtained over the past few years from neutron scattering experiments.
It is thus reasonable to ask whether these stripes are relevant for the 
mechanism of high temperature superconductivity.\cite{spingap}  Typically, in 
conventional
superconductors, the superconducting gap and transition temperature,
$T_c$, are small because they involve pairing of {\it charged particles}, 
{\it i.e.} electrons.  However, in one dimension, because of the remarkable 
fact that the low energy excitations are independent spin and charge
collective modes, superconducting pairing 
involves only the (neutral) spin degrees of freedom,  and
hence the gap can be large.\cite{spingap}  This would be a good starting
point for a mechanism of {\hty} except for the fact 
\cite{3d} that higher dimensional interactions typically lead to CDW order 
rather than superconductivity because  
the CDW susceptibility is the more divergent (see Eq. (\ref{eq:chi})) and
moreover  the Coulomb interaction between charge density fluctuations on nearby 
stripes (which promotes CDW order) is typically larger than the inter-stripe 
Josephson coupling (which promotes superconducting order). On the other hand, 
as we show below, transverse stripe fluctuations have an important
effect on the  competition between superconducting and CDW order.
Of course, such fluctuations are relatively unimportant in conventional 
quasi one-dimensional solids because of the
large mass of the constituent molecules. But the 
zero-point energy $\hbar \bar \omega$ of the transverse stripe fluctuations
becomes a significant energy scale 
in a stripe phase of a doped antiferromagnet which arises from a purely 
electronic correlation effect.

Our principal new conclusions are that, to all orders of perturbation theory
in powers of the Coulomb interaction $V$ between stripes: (i) The phase
locking between the CDW fluctuations on neighbouring stripes is
entirely eliminated by the transverse fluctions, leaving the charge
motion liquid-like along the stripe;  in other words, there exists a
stable zero temperature liquid crystalline  (quantum smectic) phase.
(ii) The Josephson coupling ${\cal J}$ between stripes is greatly enhanced by
the same transverse fluctuations. For $K_c > 1/2$, or more generally
for large enough ${\cal J}$, the ground state in this
phase is always globally superconducting.  Because of the broken
rotational symmetry of this phase, all that can be said about the
symmetry of the superconducting state is that it is singlet; its symmetry 
is necessarily a mixture of ``s-wave'' and ``d-wave''.
Conversely, for $K_c<1/2$ and
small enough ${\cal J}$, the ground state is ``metallic'' in the sense that
it has gapless charge excitations unrelated to any broken symmetry.
While such quantum critical phases are common in one dimension\cite{emery},
where they are called ``Luttinger liquids,'' we believe this is
the first theoretically well justified example in two dimensions. 

We begin with a simple model of a two-dimensional array of stripes
along the $x$ direction. 
In the ordered state, we can can safely ignore dislocations and
overhangs.  This allows us to introduce a coordinate system in which points on 
the stripes 
are labelled by a stripe number, $j$, and by a position $x$ along the stripe 
direction. 
Then the  stripe configuration is described by the transverse displacement in 
the
$y$-direction,
$Y_j(x)$, of the $j^{th}$ stripe at position $x$.       (See Fig.1.)  Because of 
the
spin gap, the only other low energy degrees of freedom involve fluctuations of 
the charge
density, $\rho_j(x)$, on each stripe,
\ba
\rho_j(x)=&& \bar \rho+\rho_0\cos[\sqrt{2\pi}\phi_j + 2k_FL_j(x)], \\
L_j(x) = && \int_0^x dx^{\prime}\sqrt{1+(\partial_{x^{\prime}}Y_j)^2} +L_j(0)
\label{eq:L}
\ea
where $\phi_j(x)$ defines the phase of the CDW with wave vector 
equal to twice the Fermi wave vector $k_F$ and 
$L_j(x)$ is the arc-length along stripe $j$.
The quantum dynamics of this system is equivalent to a theory of the 
longitudinal
($\phi_j$) and transverse ($Y_j$) vibrations of coupled
elastic strings. Technically, this defines the fixed point
Hamiltonian for the smectic phase.  (A substrate
potential would inevitably lock the smectic phase
to a wave vector commensurate with the underlying crystal lattice, and hence the
transverse modes would be gapped.\cite{pokrovsky})  

\begin{figure}
\begin{center}
\leavevmode
\epsfig{file=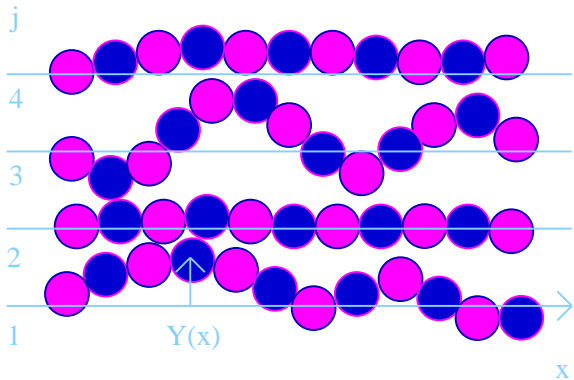,width=3in}
\end{center}
\caption
{Schematic representation of a smectic stripe phase. The coloured circles
represent periodic stuctures along the stripes, which are forced out of phase  
by the transverse fluctuations}

\label{fig1}
\end{figure}

The coupling between the  CDW's on neighbouring stripes is of the 
form
\be
H_{c}= \sum_j\int dx  \\ V(\Delta_j Y) 
\cos[\sqrt{2\pi}(\Delta_j\phi) -
2k_F(\Delta_jL)] 
\ee
plus higher harmonics. Here $L_j$ is the arc-length
defined in Eq. (\ref{eq:L}), $\Delta_j F \equiv F_{j+1}-F_j$, and the function
$V[\Delta_j Y]$ reflects the fact that CDW's on adjacent 
stripes are more
strongly coupled where the stripes are close together than where they are far
apart. When this coupling is strong, it will drive the
system into a fully-crystalline state.
Finally, there is a term in the Hamiltonian representing the 
Josephson tunnelling of (superconducting) pairs of electrons between stripes. 
The tunnelling
matrix element
\be
{\cal J}(Y) \approx {\cal J}_0\exp[\alpha Y]
\ee
depends roughly exponentially on the local spacing of the stripes.
The
fact that superconductivity is a $k=0$ order implies that the Josephson coupling
does not depend on the arc length $L_j$, and hence it is not affected by the 
geometry of the stripes.

With this background, it is possible to state our central point that,
to all orders in
perturbation theory in powers of $V$, all terms 
that are not
invariant under the transformation $\phi_j(x,\tau)\rightarrow
\phi_j(x,\tau)+\delta_j$
for arbitrary $\delta_j$ are non-vanishing only near the ``surface,''
so in the thermodynamic limit there is no locking of the phase of the
CDW fluctuations on neighbouring stripes.  (Technically, this proves that the 
fixed
point Hamiltonian is perturbatively stable.)  The physical origin of this effect 
is easily
understood.  The difference in arc lengths, $\Delta_j L = L_{j+1}(x)-L_j(x)$, is 
a sum of
contributions of random sign, and more or less independently distributed along 
the distance
$|x|$.  For this reason, $\Delta_j L$ (and the dephasing) grow with increasing
$|x|$ as in a random walk, {\it i.e.} $|\Delta_j L|^2 \sim {D|x|}$, where $D$ is 
a quantum
diffusion constant.   

This result may be obtained formally \cite{kfe} by integrating out the stripe
fluctuations ($Y$) perturbatively in powers of
$V$ and, subsequently, ${\cal J}$.  
To first order in $V$, the effective interaction between the CDW's on
neighbouring stripes, $V^{(1)}(\phi_1-\phi_2)$, is given by the expression
\be
V^{(1)}=  \langle V (\Delta Y) \cos[\sqrt{2\pi}(\Delta \phi)-2k_F(\Delta 
L)]\rangle,
\ee
where $< \ \ >$ implies averaging over transverse stripe fluctuations.  
To lowest order in a cumulant expansion 
\ba
V^{(1)}=  && \tilde V \cos[\sqrt{2\pi}(\Delta \phi)], \nonumber \\
\tilde V = && \langle V(\Delta Y)\rangle \exp \{-(2k_F^2)\langle [\Delta 
L]^2\rangle \}\nonumber \\
\approx && \langle V(\Delta Y)\rangle \exp[-(2k_F^2D)|x|]. 
\ea
This expression, which can readily be extended to higher order in perturbation 
theory
and higher order in the cumulant expansion,  captures the essential general 
point of physics-
that the coupling between CDW's vanishes very rapidly except in a region of 
width
$\sim (2k_F^2D)^{-1}$  at the ends of the stripes and hence can be ignored
in the thermodynamic limit. \cite{tech}

It is interesting to note that the quantum problem may be reformulated as a 
classical 
theory in space-time to bring out the close analogy with a well established 
phase of
conventional liquid crystals, the three-dimensional hexatic smectic B 
phase.\cite{rjb}  In the
space-time representation, the world sheets of the stripes can be regarded as 
classical
fluctuating membranes and the CDW fields 
 are analogous to a  two-dimensional hexatic phase living on the membrane.
Despite the fact that the power-law order in the plane of each
``membrane'' is modulated only in the space direction, whereas the classical
hexatic has a triangular lattice form, this analogy assures us that we have 
not omitted any important interactions from our analysis.

The effect of Josephson coupling between stripes may be analysed in the same
way.   To first order in ${\cal J}$,  the effective
action is proportional to
\be
<{\cal J}>\approx {\cal J}_0\exp\bigg\{ (\alpha^2/2)<[\Delta_jY]^2>\bigg \}. 
\ee
Notice that 
the superconducting coupling is 
strongly enhanced by the transverse stripe fluctuations.  ( Of course, there is 
a similar
enhancement of the CDW coupling, $V$, but it is overwhelmed by the dephasing 
effect.)
Physically, this enhancement reflects the fact that the mean value of  ${\cal 
J}$  is
dominated by regions where neighbouring stripes come close together so that 
$<{\cal J}>$ 
is very much larger than the median.
From Eq. (1) it can be seen that, when $K_c >1/2$, the pair susceptibility
on an individual stripe diverges as $T \rightarrow 0$ and hence  for non-zero
${\cal J}$,
the smectic phase is always globally superconducting below a finite
(Kosterlitz-Thouless) ordering temperature, $T_c\sim (\Delta_s{\cal 
J})^{K_c/(2K_c-1)}$, while
for $K_c < 1/2$ and $<{\cal  J}>$
sufficiently small, the system remains a (quantum critical) non-Fermi liquid all 
the way to $T=0$.
 
To complete the physical picture of the quantum smectic, we
construct a global phase diagram, shown schematically in 
Fig. 2, by considering the
possible zero and finite temperature phase transitions from the smectic
state to states with other symmetries.  
This can be done, to a large extent, on the basis of
general considerations of symmetry and by analogy with the phase diagram of 
conventional liquid
crystals, and the argument relies on nothing more than the existence (and 
electronic
character) of the quantum smectic phase.

\begin{figure}
\begin{center}
\leavevmode
\epsfig{file=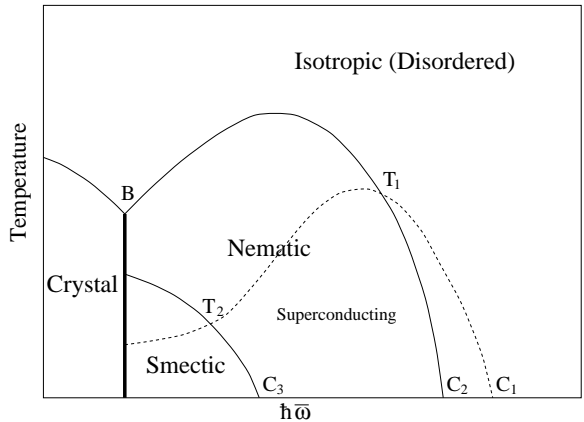,width=3in}
\end{center}
\caption
{Schematic phase diagram for $K_c>1/2$.  Here, $T$ is the temperature, 
and $\hbar \bar \omega$ is a measure of the magnitude of the transverse 
zero-point fluctuations of the stripes.  Thin lines represent continuous 
transitions and the thick line is a first order transition.  The dashed line 
is the superconducting $T_c$.  The symbols ``B'', ``C'', and ``T'' label,  
respectively, bicritical, quantum-critical, and tetracritical points.  
Depending on microscopic details, the positions of C$_1$ and C$_2$ could
be interchanged.}
\label{fig2}
\end{figure}

First consider the $T=0$ axis of the figure, in which the phases are
studied as a function of $\hbar \bar \omega$:
  
{\noindent \it i)} To the
left, as the system becomes progressively more ``classical'', {\it i.e.} for
$\hbar \bar \omega/V$ small enough, it is clear that there is a phase
transition to a crystalline state, in which the CDW order on
neighbouring stripes phase locks, the transverse stripe fluctuations
become the phonons of a fully-ordered crystal, superconducting order is
destroyed, and the system becomes globally insulating.  
This transition is typically first order.

{\noindent \it ii)} To the right, as the system becomes more quantum and, in 
particular,
when the rms magnitude of the
transverse flucuations of the stripes becomes comparable to their spacing, we 
expect
a T = 0 transition to a quantum nematic phase in which there is no broken
translational symmetry, but lattice rotational symmetry
is spontaneously broken, {\it i.e.} there are
oriented but positionally disordered stripes. We generally expect this
transition to be continuous, as shown;  this implies that, in the case
$K_c >1/2$, the superconducting order must continue across the smectic to 
nematic 
phase
boundary, and in the case $K_c < 1/2$, the Luttinger liquid behavior must 
similarly
persist across the phase boundary.  

{\noindent \it iii)}  At still larger
$\hbar \bar \omega/V$, there must be a transition to an isotropic phase. 
Landau theory suggests that the nematic to isotropic 
transition should be continuous in two spatial dimensions, although it is first 
order in three.

{\noindent \it iv)} For the case $K_c>1/2$, there are two possible scenarios for 
the
termination of the  high temperature superconducting 
order with increasing $\hbar \bar \omega$:  If the nematic region of the phase 
diagram is narrow, so that  significant
local stripe correlations survive into the isotropic phase, then one can 
imagine that the superconducting state survives until some larger value of 
$\hbar \bar \omega$, as
shown in the figure;  in this case, the superconducting state will have a pure 
symmetry
(``s'' or ``d'') where it extends into the isotropic phase.  Otherwise, the 
high temperature superconducting phase could terminate at a critical point 
within
the nematic phase.  In either case, beyond this point, the ground state is
an anisotropic Fermi liquid (similar to a conventional metal) or, if there 
remain
sufficient residual interactions, a low temperature superconductor.  
By the same logic, when the
smectic phase has a Luttinger liquid rather than a superconducting
ground state ($K_c < 1/2$), there
must be an additional zero temperature phase transition (in place of C$_1$) in 
either
the nematic or isotropic regions of the phase diagram, beyond which the system
becomes a Fermi liquid.

It is straightforward to extend this picture to $T \neq 0$. 
In isotropic two dimensional systems, at low temperature, the long-range stripe
positional order gives way to power-law order, although there is true, 
long-range
orientational (nematic) order \cite{pokrovsky}.  Ultimately, at high enough
temperature, there must be a transition to an isotropic (symmetric) phase. 
While there is the logical possibility of a direct, first order smectic to
isotropic phase transition, we consider it more likely, as at zero temperature,
that the low temperature order is destroyed in a sequence of two transitions:  
first, a
dislocation unbinding transition to a nematic phase with
short-range positional and power-law orientational order, and then a second 
transition to
the isotropic state.  

Modulo the choices among possible scenarios described above, the topology of the 
phase diagram is constrained to be that shown in Fig. 2 for the case $K_c > 
1/2$.  We have
shown the superconducting $T_c$ rising with $\hbar \bar \omega$ through the 
smectic and nematic
phases, reflecting the enhancement of the Josephson coupling, ${\cal J}$, by 
transverse
stripe fluctuations;  we show it dropping at larger $\hbar \bar \omega$ 
following the isotropic to nematic phase boundary,
since we expect that the stripes lose their local integrity far into the
isotropic phase.    A further subtlety is that both the
crystalline and smectic regions are actually a series of commensurate phases  at 
$T=0$, and a
complicated pattern of commensurate and incommensurate phases for $T\neq 0$.  
While the
commensurate smectic has true positional long-range order, the incommensurate 
smectic 
will have only power-law order.

We conclude the theoretical discussion with a few remarks. 
Crystals do not have the full rotational symmetry of free space, so the 
stripes will tend to align along a symmetry axis of the underlying crystal.
Then the smectic will acquire a finite transverse elastic modulus, rather than 
the
anomalous ``splay elastic'' constant\cite{liqcrys} of a classical
smectic.  Similarly, the incommensurate smectic and  
nematic phases will  have long-range order rather than power law behaviour.  A 
two-fold
symmetric crystal field is symmetry breaking, and changes the nematic to
isotropic phase transition into a crossover;  if the field is small, the 
crossover remains
sharp.  A four-fold or six-fold rotational symmetry will preserve this 
transition, but change it
to the Ising  or three-state-Potts universality class, respectively.\cite{HN}
The analysis of this paper can be applied to systems with low energy spin 
degrees of freedom by considering the most general model of the one 
dimensional electron gas with or without a charge gap\cite{emery,landau}.
Also, it is intuitively clear that our analysis is relevant for sufficiently
anisotropic systems, since strong three dimensional 
effects tend to make the stripes more 
rigid and in consequence will help CDW formation.

What does this have to do with the cuprate superconductors?  Tranquada {\it et 
al.} \cite{tranq}
have observed static  peaks in the spin and charge structure factors of  
{\LNSCO},
corresponding to incommensurate stripe order, with the stripes along the CuO 
direction. 
In this material, each CuO$_2$ plane has a two-fold symmetry axis that rotates
through 90$^{\circ}$ from plane to plane to give a tetragonal structure with two
inequivalent CuO$_2$ planes. The peaks have a finite width, which is consistent
with a nematic stripe phase in an orienting potential. However, the
power-law stripe order of a smectic phase provides a possible alternative 
explanation of the
observations. We feel that it is the two-fold lattice potential that drives the 
material either
into or close to the smectic phase, and freezes the dynamics. In our opinion, 
these experiments constitute strong evidence that {\LNSCO} has a low 
temperature electronic liquid crystal phase. In {\LSCO} there are similar
incommensurate peaks in the magnetic neutron scattering  factor at about the 
same position in
$\vec {k}$-space, but they are inelastic,\cite{yamada} {\it i.e.} there are
dynamically-fluctuating analogues of the stripe phases seen in {\LNSCO}. In this 
case the
structure is orthorhombic, and two-fold lattice potential is, itself, dynamical.  
Neutron scattering experiments on underdoped {\YBCO}, 
which is orthorhombic, also have found dynamical
incommensurate peaks, \cite{mook} corresponding to dynamical stripes parallel to 
the diagonal
of the CuO$_2$ unit cell. (Note that while, in general, the zero-point kinetic 
energy of 
fermions increases with increasing density,  the relationship between
$\hbar \bar \omega$ and the concentration of doped holes may be 
complicated in these materials.)

In underdoped {\hts}, two crossover lines  have been identified\cite{crossover} 
at which there
is a rather rapid change in certain physical properties, but apparently no phase 
transition. The
upper crossover is marked by a drop in the magnetic susceptibility, and the 
development of a
pseudogap in the
$c$-axis optical conductivity with a transfer of spectral weight to high 
frequencies. Both
of these features could be signatures of an isotropic-nematic transition, 
rounded by a two-fold orienting field.   It has been suggested by several 
authors\cite{romans}
that the unusual normal-state above the upper crossover temperature
is controlled by a zero-temperature quantum critical
point;  we note that the end of the  isotropic-nematic
phase boundary, C$_2$ in Fig.2,  is just such a point.  

In this paper we have focussed on doped antiferromagnetic insulators. However
other doped insulators or semiconductors could have liquid crystal phases,
especially if the band structure is anisotropic. One interesting possibility
is that, at moderately low density, a Wigner crystal may be replaced by a liquid 
crystal.


\begin{references}
\bibitem{active}   Examples of such problems include the one dimensional 
electron gas interacting with phonons, an array of Kondo or other dynamical 
impurities, and a spin chain, which may or may not be gapped.
\bibitem{tech} A technical point is that the dominant contributions to $D$ 
come from short wavelength transverse stripe fluctuations.
\end{references}
\end{document}